\documentclass[aps,prl,twocolumn,epsfig,floats]{revtex4}
\usepackage{graphicx,epsf,natbib}
\def\be{\begin{equation}}
\def\ee{\end{equation}}
\def\bea{\begin{eqnarray}}
\def\eea{\end{eqnarray}}

\begin{document}


\title{Topological Excitations near the Local Critical Point in the Dissipative 2D XY model}

\author{Vivek Aji and C. M. Varma}
\address{Physics Department, University of California,
Riverside, CA 92507}
\begin{abstract}
The dissipative XY model in two spatial dimensions belongs to a new universality class of quantum critical phenomena with the remarkable property of the decoupling of the critical fluctuations in space and time. We have shown earlier that the quantum critical point is driven by proliferation in time of topological configurations that we termed {\it warps}.  We show here that a warp may be regarded as a configuration of a monopoles surrounded symmetrically by anti-monopoles so that the total charge of the configuration is zero. Therefore the interaction with other warps is local in space. They however interact with other warps at the same spatial point logarithmically in time. As a function of dissipation warps unbind leading to a quantum phase transition. The critical fluctuations are momentum independent but have power law correlations in time. 
\end{abstract}
\maketitle

One of the important developments in condensed matter physics of the last quarter century is the realization that there exists a class of quantum-critical phenomena that are not simply extensions \cite{moriya1, beal-monod, hertz, millis} of the well understood dynamical critical phenomena near classical critical points. In the usual extensions, an essential feature of classical critical fluctuations, that the correlation length in the temporal-direction is a power $z$ of the correlation length in the spatial direction, is maintained. It was observed by contrast, in one part of the phase diagram of  the normal state of the high temperature cuprate superconductors, that the properties could be understood only by quantum-critical fluctuations which are local spatially but with a simple ($\propto 1/t$) power law correlation in time \cite{mfl}. 

Subsequently similar local criticality has been adduced from measurements of the correlation functions for some heavy-fermion compounds near their antiferromagnetic quantum-critical point  \cite{lohneysen, schroder, schroder1, si}. It has also been suggested that disordered superconducting thin films \cite{ORR, haviland}, and superconducting wires \cite{bezryadin, lau} have local  quantum-criticality at the superconducting to insulating transition \cite{SC, SC1, MPAF, MPAF1}. Local phase slips have been shown to play a crucial role in dissipation driven transitions in one dimensional superconducting wires \cite{GIL}. The wire can be thought of as a linear chain of coupled superconducting grains. Each event in space-time creates a dipole consisting of a phase slip in one junction and an anti-phase slip in a neighboring junction. 

 In all the physical situations mentioned above, one may identify regimes, in parameters like temperature, doping, pressure, thickness and magnetic field or charge density, where the global properties are  dominated by local degrees of freedom. The absorptive part of the fluctuation spectrum of these degrees of freedom displays $\omega/T$ scaling.  This is simply the thermally weighted Fourier transform of $1/t$ correlations. There is no evidence of a spatial correlation length which is related to such a temporal correlation.  

For the dissipative 2DXY model, a variational and  weak-coupling calculations suggested that the quantum-criticality is local \cite{SC,SC1}. To compute the correlation functions, a non-perturbative treatment is necessary. Recently, such a theory of this and some related models has been developed \cite{aji1, aji2}. The quantum criticality of cuprates belongs to this universality class\cite{aji1,aji2,aji3,as1}. The principal accomplishment is to show that spatial and temporal degrees of freedom of fluctuations exactly decouple in the partition function because the fluctuations can be represented in terms of orthogonal topological defects - vortices which are shown to interact locally in time and logarithmically in space and a new class of defects {\it warps}, which interact locally in space and logarithmically in time.

In this paper we resolve some remaining questions about the discrete nature of the warps and explain why the warps interact locally in space. We do this  by giving an explicit representation for them in terms of configuration of phase slips for the 2d dissipative xy model and give a physical picture as to why they are orthogonal to the vortices. This is the only two-dimensional model for which local quantum-criticality and $\omega/T$ scaling have been explicitly proven. As such, the general understanding and technical procedure may be important for other problems where similar criticality appears to govern.

This paper is organized as follows. In the next section, we summarize briefly the transformations \cite{aji1, aji2} leading to the Action which displays decoupling of space and time variables. In the subsequent sections, we introduce Vortices and Warps, give their physical description in terms of configurations of phase slips, and  show that the Action in terms of them is asymptotically identical to the Action derived earlier.

\section{Dissipative 2DXY model and its dual representation} 

\begin{figure}
  \begin{center}
  \includegraphics[width=0.8\columnwidth]{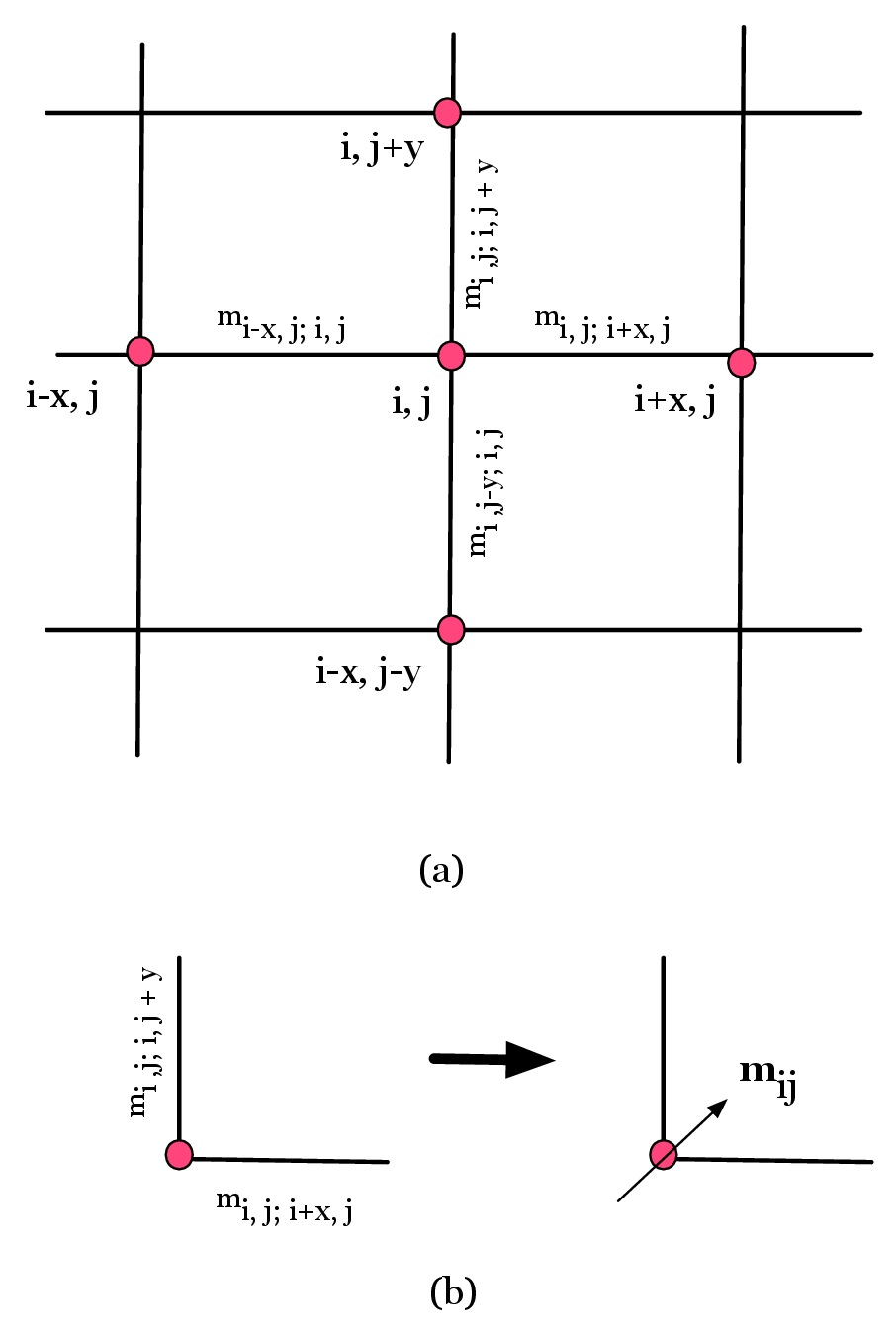}
  \caption{The directed link variables are labelled as shown in the (a). (b )We define a two component vector living on the sites of the original lattice whose components are the two directed links variables: $\textbf{m}$ = $\left( m_{i,j;i+x,j}, m_{i,j;i,j+x}\right)$}
  \label{fig:2dd}
  \end{center}
\end{figure}

The classical 2d xy model consists of  $U(1)$ degrees of freedom,
represented by an angle $\theta$, living on the sites $(ij)$ of a regular
lattice, assumed here to be a square lattice see Fig. (\ref{fig:2dd}), with a nearest
neighbor interaction of the form

\begin{equation}
H = J\sum_{\left\langle ij,kl \right\rangle}\left[1 -
\cos\left(\theta_{ij} - \theta_{kl}\right)\right].
\end{equation}

\noindent  Since a continuous symmetry cannot be spontaneously broken in two dimensions \cite{MW,PCH}, this model does not support a long range ordered phase.
Nevertheless a phase transition does occur at finite temperature where the correlation function of the order parameter $e^{\imath\theta}$ changes from exponential to power law. This is the
Kosterlitz-Thouless-Berezinskii transition \cite{BER, KT}.  The quantum dissipative generalization of the model includes two dynamical terms and is given by

\begin{widetext}
\begin{eqnarray}\label{q2dxy}
Z  &=& \int D \theta_{i}\left(\tau\right) \exp\left[-\int_{0}^{\beta}
d\tau \left(\sum_{i}{C \over
{2}}\left(\partial_{\tau}\theta_{ij}^{2}\right) - J\sum_{\left\langle
ij,kl\right\rangle}\cos\left(\theta_{ij}-\theta_{kl}\right)\right)+ S_{diss}\right]\\ \nonumber
S_{diss} &=& \int_{-\infty}^{\infty}d\tau \int_{0}^{\beta}d\tau
'\sum_{\left\langle ij,kl
\right\rangle}\alpha\left({\left(\theta_{ij}-\theta_{kl}\right)\left(\tau\right)-\left(\theta_{ij}-\theta_{kl}\right)\left(\tau
'\right)\over {\tau - \tau '}}\right)^{2},
\end{eqnarray}.
\end{widetext}

\noindent where $C$ is the capacitance and $\alpha=R_{Q}/R$ where $R_{Q}=h/4e^{2}$.

The physics of this phase transition is better understood in terms of the topological defects of the system. To do so we follow the
standard procedure of using the Villain transform and integrating out the phase degrees of freedom \cite{aji2}. The Villain transform involves expanding the periodic
function in terms of a periodic Gaussian

\begin{widetext}
\begin{equation}\label{vill}
\exp\left[-\beta J\sum_{\left\langle ij,kl \right\rangle}\left[1 -
\cos\left(\theta_{ij} - \theta_{kl}\right)\right]\right] \approx
\sum_{m_{ij;kl}}\exp\left[-\beta J\sum_{\left\langle ij,kl
\right\rangle}\left(\theta_{ij}-\theta_{kl}-2\pi
m_{ij;kl}\right)^{2}/2\right],
\end{equation}
\end{widetext}

\noindent where $m_{ij;kl}$ are integers that live on the links of the
original square lattice. We can combine
the two link variables $m_{i,j;i+1,j}$ and $m_{i,j;ij+1}$ into one
two component vector $\textbf{m}_{i,j}$ that lives on the site $\{i,j\}$
of the lattice (see fig.\ref{fig:2dd}).  We expand the quadratic term and transform to Fourier
space. Keeping the leading quadratic term
$\theta_{i,j}-\theta_{i+1,j}\approx$ $- a\nabla_{x}\theta_{xy}$,
where a is the lattice constant, $x = a i$ and $y = a j$, we get

\begin{eqnarray}
\left(\theta_{xy}-\theta_{x+1,y}-2\pi m_{x,y}^{x}\right)^{2}\approx
a^{2}\nabla^{2}_{x}\theta_{xy} &+& 4\pi
a\nabla_{x}\theta_{xy}m_{x,y}^{x} \nonumber \\&+& 4\pi^{2}m_{x,y}^{x
2},
\end{eqnarray}

\noindent where $m^{x}_{x,y}$ is the $x$ component of the vector
field given by the integer $m_{x,y;x+1,y}$. In the absence of dissipation, there
is competition between the kinetic energy and the potential energy
terms, the former minimized by a state where $\theta_{ij}$ is disordered
stabilizing an insulating phase while the latter minimized by a
fixed value of $\theta_{ij}$ stabilizing a superconductor. Since $\theta$'s are Bosonic degrees of freedom,  we need to impose the boundary condition that $\theta_{ij}\left(\beta\right) = \theta_{ij} (0)$. The periodicity in the imaginary time direction and the compactness of the field $\theta_{ij}$ implies that there is an additional degree of freedom that has to be accounted for which is the winding number.At $T=0$, the (imaginary) time direction becomes infinite in extent, and the non-dissipative model is in the 3DXY universality class. First we discretize the imaginary time direction in units of $\Delta\tau$ and work on a three dimensional lattice. Introducing the variables ${\bf m}$ which only live on the spatial links, the action  is \cite{aji1, aji2}

\begin{widetext}
\begin{eqnarray}\label{dissvrt}
Z  = \sum_{\textbf{m}}\exp\left[\sum_{\textbf{k},\omega}-4\pi^{2} J
\frac{Jc\left|\textbf{k}\times\textbf{m}\right|^{2}} {\left(C/c\right)\omega_n^{2}+Jc k^{2}+\alpha\left|\omega_{n}\right|k^{2}}- 4\pi^{2}J{\omega_{n}^{2}\textbf{m}\cdot\textbf{m}\left(C/c+\alpha
k^{2}/\left|\omega_{n}\right|\right)\over {\left(C/c\right)\omega_n^{2}+Jc k^{2}+\alpha\left|\omega_{n}\right|k^{2}}}\right],
\end{eqnarray}
\end{widetext}

\noindent where $c = a/\Delta\tau$, $J\rightarrow Ja^{2}\Delta\tau$, $C\rightarrow Ca^{2}/\Delta\tau$ and $\textbf{m}\rightarrow \textbf{m}/a$. We have also redefined $\alpha\rightarrow \alpha a^{3}$. The last term in the denominator is unimportant and may be dropped. The action in Eqn.\ref{dissvrt} has two possible phase transitions, depending on whether  the capacitance $C$ or the dissipation term $\propto \alpha$  in the numerator of the second term in Eq. (\ref{dissvrt})
dominates in long wavelength and low frequency limit. The former corresponds a critical point with the dynamic critical $z=1$, i.e. we recover the loop gas model which belongs to the 3d xy universality class. The latter corresponds, as we will show to $z = \infty$, i.e.  a fixed point with local criticality. The two
terms have the same scaling form for $z=2$.  

Suppose we separate ${\bf m}$ into the usual transverse part in terms of a vortex field $ \rho_v ({\bf k}, \omega_n)$
\bea
\label{rhov}
i {\bf k}\times {\bf m}_t({\bf k}, \omega_n) = \rho_v ({\bf k}, \omega_n)
\eea
and a {\it longitudinal} part through introducing the {\it warps} $ \rho_w ({\bf k}, \omega)$
\bea
\label{rhow}
 \omega_n  {\bf m}_{\ell}({\bf k}, \omega_n) =  c \hat{k}\rho_w ({\bf k}, \omega_n).
\eea
Quite miraculously the partition function in Eq. (\ref{dissvrt}) can be written exactly as 
\begin{eqnarray}
\label{separation}
Z &=&  \sum_{\rho_{v},\rho_{w}} \exp\left[ \sum_{{\bf k}, \omega_n}\left\{
{J\over {k^{2}}}\left|\rho_{v} (\textbf{k}\omega_{n})\right|^{2}\right.\right.\\
\nonumber &-&{\alpha\over
{4\pi\left|\omega\right|}}\left|\rho_{w}(\textbf{k}\omega_{n})\right|^{2}
\\ \nonumber  &-& \left.\left.
G\left(\textbf{k},\omega_{n}\right) \left(J J_{t} - {\alpha
J_{t}\left|\omega_{n}\right|\over {4\pi c}} - {\alpha^{2}k^{2}\over
{16\pi^{2}}}\right) 
\left|\rho_{w}(\textbf{k}\omega_{n})\right|^{2}\right\}\right]
\end{eqnarray}

\noindent where

\begin{equation}
G\left(\textbf{k},\omega_{n}\right) =  {1\over{J c k^{2} +
\left(C/c\right)\omega_{n}^{2}+\alpha\left|\omega_{n}\right|k^{2}}}
\end{equation}
The first terms describes vortices interacting logarithmically in space with no retardation in time, the second describes warps interacting logarithmically in time with local interactions and the last term is irrelevant near the critical point. The partition function near criticality can then be easily be evaluated. This is the procedure followed in Ref. (\onlinecite{aji1, aji2}) to evaluate the correlation functions near criticality.

Two questions immediately arise. (1)  Is the discreteness of the variables ${\bf m}_{\ell}$ being respected in this procedure? While this is not obvious even for a vortex in the continuum equation for vortices, Eq. (\ref{rhov}), it is well understood that such equations indeed respect the discreteness of ${\bf m}_t$ through the discreteness of $\rho_v$. A less than satisfactory answer for the discreteness of the variables ${\bf m}_{\ell}$ was provided in Ref.\onlinecite {aji2}. A clear demonstration is presented below. (2) While the locality of the interaction between warps is explicitly shown in Eq. (\ref{separation}), it is worthwhile to show this in direct physical terms. This is also accomplished below.

\section{Topological Defects}

\subsection{Vortices}

\begin{figure}
  \begin{center}
  \includegraphics[width=0.8\columnwidth]{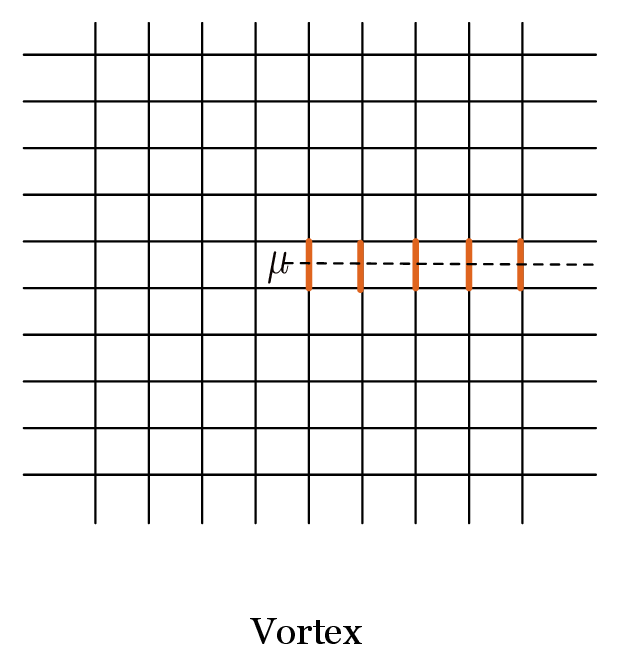}
  \caption{An isolated vortex at site $\mu$. In terms of the vector field $\textbf{m}$, the vortex is a string (dashed line) that stretches out to the system boundary. All links intersected by the string have a constant $m_{ij}$ represented by the red bonds.}
  \label{fig:vortex}
  \end{center}
\end{figure}

The 2Dxy model supports topological excitations where the vector field $\textbf{m}$ acquires nonzero curl. An isolated vortex, of unit vorticity, at site $\mu$ is represented in the fig.\ref{fig:vortex}.  In terms of the dual variables $\textbf{m}$, the vortex is a nonlocal object. A string stretching from site $\mu$ (greek alphabets are used for sites of the dual lattice) to infinity is a representation of the vortex with the link variables being finite only on those links intersected by the string. If $\left(x_{lr\mu},y_{lr\mu}\right)$ labels the lower right hand corner of the plaquette labeled $\mu$, we can represent the vortex as 

\begin{equation}
\textbf{m}_{v} \left(x, y\right) = \widehat{y} \Theta\left(x- x_{lr\mu}\right)\delta(y- y_{lr\mu})
\end{equation}

\noindent where $a$ is the lattice constant. Note that $\nabla \times \textbf{m} = \delta (\textbf{r} -\textbf{r}_{i})$. An ensemble of such strings faithfully generates all possible configurations of vortices. The vortex degrees of freedom are defined as $\nabla \times \textbf{m}\left(\textbf{r}\right) = \rho_{v}\left(\textbf{r}\right)$.

\noindent The classical partition function is completely determined by the vortex degrees of freedom. The 2DXY model is mapped to a neutral plasma of vortices (zero total vorticity). As  temperature is raised  through the critical temperature, vortices unbind and proliferate. As a result the order parameter correlation function changes from power law at low temperatures to exponential at high temperatures.

\noindent {\it Dynamics}: The quantum model includes the dynamics of the vector field $\textbf{m}$. The time evolution of the vector field results in a vortex current given by

\begin{equation}
\textbf{J}_{v} = \widehat{z} \times {d \textbf{m}\over {d \tau}}
\end{equation}

\noindent The vortex density and current satisfy the continuity equation. In the absence of dissipation, the partition function is

\begin{widetext}
\begin{equation}
\label{non-diss}
Z =  \sum_{\textbf{m}}\exp\left[\sum_{\textbf{k},\omega_{n}}-4\pi^{2} J{ J c \left|\rho_{v}\left(\textbf{k}, \omega_{n}\right)\right|^{2} + \left(C/c\right) \left|\textbf{J}_{v}\left(\textbf{k}, \omega_{n}\right)\right|^{2}\over {\left(C/c\right)\omega_n^{2}+Jc k^{2}}}\right]
\end{equation}
\end{widetext}

\noindent Eq. (\ref{non-diss}) shows the well known fact that the quantum 2DXY model without dissipation maps onto a three dimensional vector field with coulomb interaction and a constraint of no divergence (continuity equation). In addition to spatial phase slips on links this model allows for phase slip events in time.

\begin{figure}
  \begin{center}
  \includegraphics[width=0.8\columnwidth]{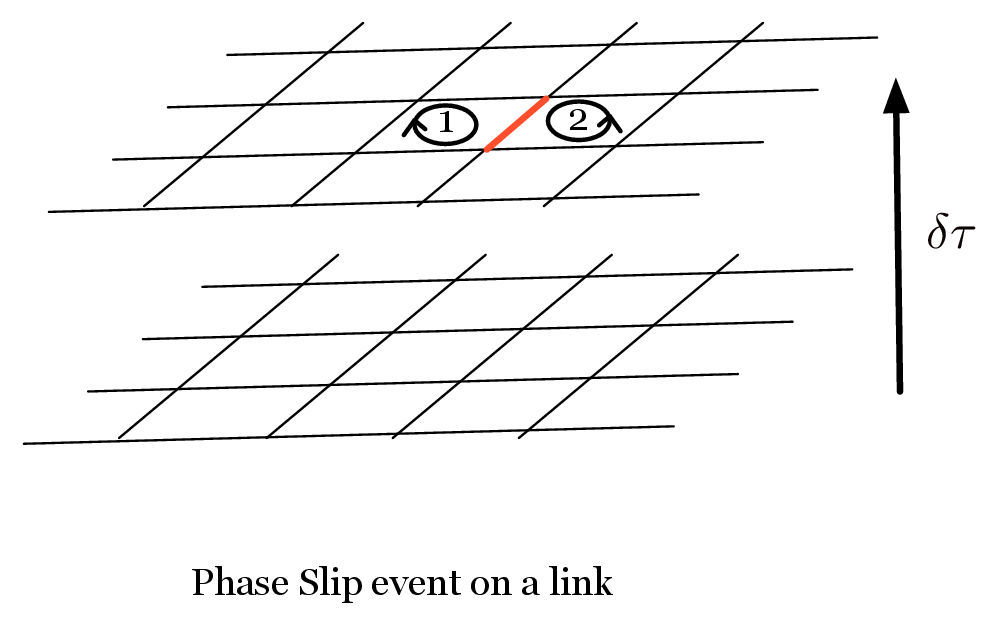}
  \caption{A phase slip between nearest neighbors in time produces a vortex and an antivortex on neighbouring plaqiettes. In time one bond has acquired a nonzero value, represented by the red bond. If the value on the bond is -1, the vorticity at site 1 is negative and on site 2 is positive.}
  \label{fig:phasesliponlink}
  \end{center}
\end{figure}

\noindent Since the current involves a change in the value of the vector field, consider an event in time where a single bond acquires a nonzero value. Such an event produces a vortex and and antivortex on neighboring plaquettes as shown in fig.\ref{fig:phasesliponlink}

\noindent Alternatively, one can interpret the event of creation of a vortex dipole as an vortex current. In time interval $\delta\tau$ a vortex moves from site 1 to site 2 in fig.\ref{fig:phasesliponlink}, leaving behind an antivortex at site 1. The total vorticity before and after the event is zero as is required by the condition of having a neutral plasma. Over time $\delta \tau$ a phase slip event occurs on the link between sites 1 and 2 at time $\tau_{i}$. If the lower end of the bond is labeled by $\textbf{r}_{xy}=\{x, y\}$, the event, in our notation, is given as

\begin{equation}
\textbf{m} = -\widehat{y}\delta\left(\textbf{r}-\textbf{r}_{xy}\right)\Theta\left(\tau-\tau_{i}\right)
\end{equation}

\noindent The vortex current generated by such an event is

\begin{equation}
\textbf{J}_{V} = \widehat{x}\delta\left(\textbf{r}-\textbf{r}_{xy}\right)\delta\left(\tau-\tau_{i}\right)
\end{equation}

\noindent Thus a phase slip on a link is equivalent to a local in space and time vortex current. 

\subsection{Local phase slips and Warps}

\begin{figure}
  \begin{center}
  \includegraphics[width=0.8\columnwidth]{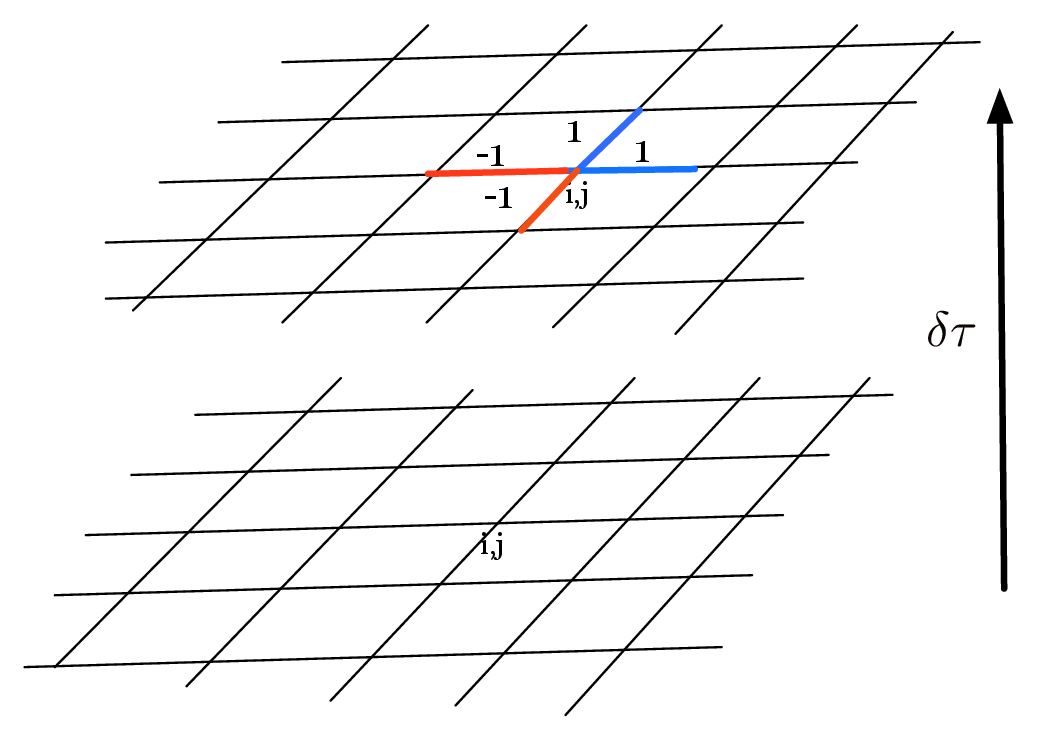}
  \caption{A phase slip event in rime results in a change in the link variables. For a change of $2\pi$ at site $\{i,j\}$, the four links connected to it acquire the values shown in the figure.}
  \label{fig:warps}
  \end{center}
\end{figure}

The periodic in time boundary condition allows for phase slip events on a site. Such events change the winding sector and dynamics of the vector field that is not captured by the vortex current. Consider the effect of a change of $2\pi$ at site $\{i, j\}$ so that $\theta_{ij}$ winds around to $\theta_{ij}+2\pi$ over a time $\delta\tau$. Then the four spatial links connected to the site $\{i, j\}$ experience phase slips. As shown in Fig. (\ref{fig:warps}), the corresponding link variables are: $m_{i,j:i+x,1}=1$, $m_{i,j:i,j+1}=1$, $m_{i,j:i-1,j}= -1$ and $m_{i,j:i,j-1}=-1$.

\noindent The phase slip event at time $\tau_{i}$ and site $\textbf{r}_{ij}$ has the following representation in terms of the vector field $\textbf{m}$:

\begin{eqnarray}\label{warpm}
\textbf{m}(\textbf{r},\tau) &=&\left[ \left(\widehat{x}+\widehat{y}\right)\delta\left(\textbf{r}-\textbf{r}_{ij}\right) - \widehat{x}\delta\left(\textbf{r}-\textbf{r}_{ij}- a \widehat{x}\right)\right. \\ \nonumber &-&\left.  \widehat{y}\delta\left(\textbf{r}-\textbf{r}_{ij}- a \widehat{y}\right)\right]\Theta\left(\tau-\tau_{i}\right) 
\end{eqnarray}

\noindent Such a vector field distribution has no curl and hence does not effect the vorticity. On the other hand the divergence is nonzero and phase slip events generate field configurations that are orthogonal to those created by vortices. For a general vector field one does expect two kinds of sources to generate an arbitrary distribution. For the 2+1 dimensional quantum model we have, besides the vortices, the additional topological entity to describe the winding number sector in time. Events that change the winding number sector, i.e. local phase slips, acts as sources for a divergence in the vector field. Just as a vortex is equivalent to an electric charge in the dual language, the sources created by phase slips can be shown to be a local distribution of monopoles ($\rho_{m}$). Given the distribution in eqn.\ref{warpm}, the corresponding configuration of monopoles, which we term the charge of the {\it phaseslip}  is

\begin{eqnarray}
\rho_{m}\left(\textbf{r}, \tau\right) &=&  \nabla\cdot\textbf{m}(\textbf{r},\tau) \\ \nonumber
&=& \left[ 4\delta\left(\textbf{r}-\textbf{r}_{ij}\right) -\delta\left(\textbf{r}-\textbf{r}_{ij}+ a \widehat{x}\right)\right. \\ \nonumber &-& \delta\left(\textbf{r}-\textbf{r}_{ij}- a \widehat{x}\right)
- \delta\left(\textbf{r}-\textbf{r}_{ij}+ a \widehat{y}\right)\\ \nonumber &-& \left. \delta\left(\textbf{r}-\textbf{r}_{ij}- a \widehat{y}\right)\right]\Theta\left(\tau-\tau_{i}\right) 
\end{eqnarray}

\begin{figure}
  \begin{center}
  \includegraphics[width=0.8\columnwidth]{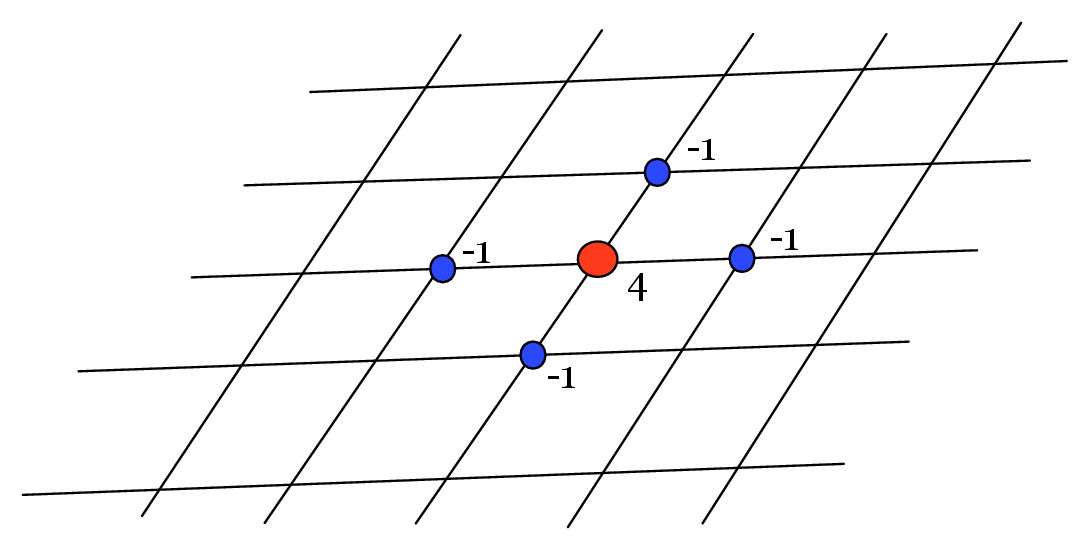}
  \caption{Warps are configurations of the field $\textbf{m}$ with finite divergence but no curl. For a warp of unit strength at site $\{i, j\}$, the divergence has a magnitude of 4 at the site and -1 at the four nearest neighbor sites. In general the magnitude of the divergence at the site will equal to the number of nearest neighbors on the lattice.}
  \label{fig:monopoles}
  \end{center}
\end{figure}

\noindent The monopole distribution equivalent to a phaseslip is shown in fig.\ref{fig:monopoles}. The total monopole charge of the configuration is zero. Since the distribution has azimuthal symmetry all harmonics are zero. This is the two dimensional lattice realization of the configuration of a charge surrounded by an equal but opposite charge distributed over a spherical shell of radius $a$ in three dimensions. The magnetic field due to the charges is confined within one unit cell around the site of the phase slip and is zero outside. Thus two phase slip events can interact only if they are at most one lattice spacing apart; the interaction is local in space. Although this is physically obvious from this discussion, we will demonstrate this explicitly in the next section. 

\noindent Phase slip events generate a local vortex current which is divergenceless but has a finite curl.

\begin{eqnarray}
J_{ps}\left(\textbf{r}, \tau\right) &=&  \left[ \left(-\widehat{x}+\widehat{y}\right)\delta\left(\textbf{r}-\textbf{r}_{ij}\right) - \widehat{y}\delta\left(\textbf{r}-\textbf{r}_{ij}- a \widehat{x}\right)\right.  \nonumber \\ &+&\left.  \widehat{x}\delta\left(\textbf{r}-\textbf{r}_{ij}- a \widehat{y}\right)\right]\delta\left(\tau-\tau_{i}\right) 
\end{eqnarray}

\noindent The continuity equation is satisfied as this current has no divergence.  

\section{Dissipation driven transition}

Any configurations of the  vector field $\textbf{m}$ can be described by an ensemble of phase slips and vortices. Since the former are divergence free and the latter curl free, we have

\begin{eqnarray}
\nabla\times \textbf{m}_{t}\left(\textbf{r}, \tau\right)  &=& \rho_{v}\left(\textbf{r}, \tau\right) \\ \nonumber
\textbf{m}_{l}\left(\textbf{r}, \tau\right)  &=&\sum_{i}\rho_{ps}^{i} \Theta\left(\tau-\tau_{i}\right) \left[ \left(\widehat{x}+\widehat{y}\right)\delta\left(\textbf{r}-\textbf{r}_{ij}\right) \right. \\ \nonumber &-&  \widehat{x}\delta\left(\textbf{r}-\textbf{r}_{ij}- a \widehat{x}\right)\\ \nonumber &-&\left.  \widehat{y}\delta\left(\textbf{r}-\textbf{r}_{ij}- a \widehat{y}\right)\right]
\end{eqnarray}

\noindent In Fourier space we get

\begin{eqnarray}\label{vecm}
\textbf{m}_{t}\left(\textbf{k}, \omega_{n}\right) &=& {\imath\widehat{z}\times\textbf{k} \over {k^{2}}}\rho_{v}\left(\textbf{k}, \omega_{n}\right) \\ \nonumber
\textbf{m}_{l}\left(\textbf{k}, \omega_{n}\right) &=&{\left(1-e^{-\imath k_{x}a}\right)\widehat{x}+\left(1-e^{-\imath k_{y}a}\right)\widehat{y}\over {\imath\omega_{n}}} \rho_{ps}\left(\textbf{k}, \omega_{n}\right)\\ \nonumber
 \rho_{ps}\left(\textbf{k}, \omega_{n}\right) &=& \sum_{i}\rho_{ps}^{i}e^{-\imath \textbf{k}\cdot\textbf{r}_{i}-\imath\omega_{n}\tau_{i}}
\end{eqnarray}

\noindent The locality in space is reflected in eqn.\ref{vecm} by the factors $\left(1-\exp\left(-\imath k_{x}a\right)\right)$ and $\left(1-\exp{\left(-\imath k_{y}a\right)}\right)$. Since $k_{x}a, k_{y}a \ll 1$, to leading order we get

\begin{equation}\label{mlwv}
\textbf{m}_{l}\left(\textbf{k}, \omega_{n}\right) \approx {a\left(k_{x} \widehat{x} + k_{y}\widehat{y}\right)\over {\omega_{n}}} \rho_{ps}\left(\textbf{k}, \omega_{n}\right)
\end{equation} 

\noindent The two components of the field $\textbf{m}_{t}$ and $\textbf{m}_{l}$ are orthogonal to each other. Comparing the longitudinal component of $\textbf{m}_{l}$ generated by a phase slip in the long wavelength limit and a warp (eqn.\ref{rhow}), we find they are proportional to each other. In particular $ka\rho_{ps} = \rho_{w}$. The partition function can be recast in terms of the charges. 

\begin{widetext}
\begin{eqnarray}\label{chgact}
Z &=& \sum_{\{\rho_{v},\rho_{ps}\}}\exp\left[\sum_{\textbf{k},\omega_{n}} -4\pi^{2} {J\over k^{2}} \left|\rho_{v}\left(\textbf{k}, \omega_{n}\right)\right|^{2} -4\pi^{2}{\alpha\over {c\left|\omega_{n}\right|}}\left(4-2\cos\left(k_{x}a\right)-2\cos\left(k_{y}a\right)\right) \left|\rho_{ps}\left(\textbf{k}, \omega_{n}\right)\right|^{2} \right. \\ \nonumber &-& \left. 4\pi^{2}{JC/c - \alpha\left|\omega_{n}\right|C/c^{2}-\alpha^{2}k^{2}/c\over {{\left(C/c\right)\omega_n^{2}+Jc k^{2}+\alpha\left|\omega_{n}\right|k^{2}}}}\left(4-2\cos\left(k_{x}a\right)-2\cos\left(k_{y}a\right)\right) \left|\rho_{ps}\left(\textbf{k}, \omega_{n}\right)\right|^{2} \right]
\end{eqnarray}
\end{widetext}

\noindent The singularities decouple space and time. Vortices interact logarithmically in space with no retardation while phase slips interact logarithmically in time. The interaction between phase slips is short range in that two phase slips interact only if they are less than a lattice spacing apart. As pointed out in the previous section, phase slips are a localized configuration of monopoles whose magnetic field is confined to one unit cell. Thus the interaction is short ranged. 

\noindent The form of the action is similar to Eq. (\ref{separation}), which was the basis for the calculations of Ref. [\onlinecite{aji1,aji2}].  The only difference is in the factor of $\left(4-2\cos\left(k_{x}a\right)-2\cos\left(k_{y}a\right)\right) $ that multiplies the terms involving warps, which comes from explicitly deriving the action on a lattice  as opposed to in a continuum in Eq. (\ref{dissvrt}). The nature of singularity is not altered. To make this point explicitly, consider the long wavelength limit of the most singular part of the action for the phase slips (second term in eqn.\ref{chgact}). For  $k_{x}a, k_{y}a \ll 1$, this term is (upto factors of $4\pi^{2}$)

\begin{eqnarray}
&&{\alpha\over {c\left|\omega_{n}\right|}}\left(4-2\cos\left(k_{x}a\right)-2\cos\left(k_{y}a\right)\right) \left|\rho_{ps}\left(\textbf{k}, \omega_{n}\right)\right|^{2}\nonumber\\  &\approx & {\alpha\over {c\left|\omega_{n}\right|}}k^{2}a^{2} \left|\rho_{ps}\left(\textbf{k}, \omega_{n}\right)\right|^{2} ={\alpha\over {c\left|\omega_{n}\right|}} \left|\rho_{w}\left(\textbf{k}, \omega_{n}\right)\right|^{2}
\end{eqnarray}

\noindent The fact that the discrete phase slips interact locally in space translates to a local interaction among warps in the long wavelength limit. Thus the critical theory are identical. At the microscopic level, the connection with phase slips guarantees that the discrete nature of the $\textbf{m}$ fields is obeyed. 

The last term in eqn.\ref{chgact} is not singular as it involves a scalar field with three dimensional Coulomb interaction. The most singular interaction among the phase slips in real space can be written as

\begin{widetext}
\begin{eqnarray}
Z_{w} &=&  \sum_{\{\rho_{ps}\}}\exp\left[\sum_{i}\int {d\tau}\int d\tau ' 4\alpha\rho_{ps}\left(\textbf{R}_{i},\tau\right)\log\left|\omega_{c}\left(\tau - \tau '\right)\right|\rho_{ps}\left(\textbf{R}_{i},\tau'\right)\right.\\ \nonumber &-& \left. \sum_{i,j}\int {d\tau}\int d\tau ' \alpha\rho_{ps}\left(\textbf{R}_{i},\tau\right)\log\left|\omega_{c}\left(\tau - \tau' \right)\right|\rho_{ps}\left(\textbf{R}_{i}+\textbf{r}_{j},\tau'\right)\right]
\end{eqnarray}
\end{widetext}

\noindent where $r_{j}$ runs over the four nearest neighbors of $i$ on the square lattice and $\omega_{c}$ is the high frequency cutoff. Furthermore the singularity at $\omega = 0$ implies that at each site the sum of all charges over time is zero. In addition to on site long range in time interaction, phase slips involve interaction among nearest neighbors. As $\alpha$ is varied there is a unbinding transition in time at $\alpha_{c}=1/4$. The phase transition is identical to the one analyzed in ref.[ \onlinecite{aji1,aji2}]. The correlation functions are local in space and power law in time.

\section{Conclusion}

The local character of the quantum phase transition of the dissipative 2DXY model is a consequence of a new topological defect, warps,  that arises due to phase slip events in time. The discrete nature of the charges is related to the periodic in time boundary condition enforced on bosonic fields in the path integral formalism. The compactness of the phase field $\theta$ allow for world lines in time that possess nontrivial winding numbers. The events in time where the winding numbers change correspond to spatial configuration of phase slips between nearest neighbor links that are orthogonal to the vortices. While in dissipation free systems such charges contribute nonsingular terms to the effective action, ohmic dissipation yields a new class of quantum critical point. 

The universality class defined by such singularities is a new direction in the study of quantum critical phenomena with potential realizations in a number of systems such as cuprate superconductors, disordered two dimensional thin films,  heavy fermion superconductors and possibly the pncitide superconductors near their antiferromagnetic quantum-critical point, as well as plateau transitions in quantum Hall effects. We speculate that these systems also belong to the universality class where the spatial and temporal correlations are decoupled due to orthogonal topological defects. The precise nature of the defects is bound to have different microscopic origin and should be the subject for future studies.

The authors wish to acknowledge David Clarke, Gil Rafael, and Kirill Shtengel for their comments and questions. VA research is supported by University of California at Riverside under the initial complement. CMV's research is partially supported by NSF grant DMR-0906530.


\begin{thebibliography}{10}

\bibitem{moriya1}
T\^{o}ru Moriya and Arisato Kawabata.
\newblock {\em Journal of the Physical Society of Japan}, 34(3):639--651, 1973.

\bibitem{beal-monod}
M.~T. B\'eal-Monod and Kazumi Maki.
\newblock {\em Phys. Rev. Lett.}, 34(23):1461--1464, 1975.

\bibitem{hertz}
John~A. Hertz.
\newblock {\em Phys. Rev. B}, 14(3):1165--1184, Aug 1976.

\bibitem{millis}
A.~J. Millis.
\newblock {\em Phys. Rev. B}, 48(10):7183--7196, Sep 1993.

\bibitem{mfl}
C.~M. Varma, P.~B. Littlewood, S.~Schmitt-Rink, E.~Abrahams, and A.~E.
  Ruckenstein.
\newblock {\em Phys. Rev. Lett.}, 63(18):1996--1999, Oct 1989.

\bibitem{lohneysen}
Hilbert v.~L\"{o}hneysen, Achim Rosch, Matthias Vojta, and Peter W\"{o}lfle.
\newblock {\em Reviews of Modern Physics}, 79(3):1015, 2007.

\bibitem{schroder}
A.~Schroder, G.~Aeppli, R.~Coldea, M.~Adams, O.~Stockert, H.v. Lohneyesen,
  E.~Bucher, R.~Ramazashvili, and P.~Coleman.
\newblock {\em Nature}, 407:351--355, 2000.

\bibitem{schroder1}
A.~Schr\"oder, G.~Aeppli, E.~Bucher, R.~Ramazashvili, and P.~Coleman.
\newblock {\em Phys. Rev. Lett.}, 80(25):5623--5626, Jun 1998.

\bibitem{si}
Q.~Si, S.~Rabello, K.~Ingersent, and J.~Lleweilun Smith.
\newblock {\em Nature}, 416:610, 2002.

\bibitem{ORR}
B.~G. Orr, H.~M. Jaeger, A.~M. Goldman, and C.~G. Kuper.
\newblock {\em Phys. Rev. Lett.}, 56(4):378--381, Jan 1986.

\bibitem{haviland}
D.~B. Haviland, Y.~Liu, and A.~M. Goldman.
\newblock {\em Phys. Rev. Lett.}, 62(18):2180--2183, May 1989.

\bibitem{bezryadin}
A.~Berzryadin, C.N. Lau, and M.~Tinkham.
\newblock {\em Nature}, 404(6781):971, 2000.

\bibitem{lau}
C.~N. Lau, N.~Markovic, M.~Bockrath, A.~Bezryadin, and M.~Tinkham.
\newblock {\em Phys. Rev. Lett.}, 87(21):217003, Nov 2001.

\bibitem{SC}
Sudip Chakravarty, Gert-Ludwig Ingold, Steven Kivelson, and Alan Luther.
\newblock {\em Phys. Rev. Lett.}, 56(21):2303--2306, May 1986.

\bibitem{SC1}
Sudip Chakravarty, Gert-Ludwig Ingold, Steven Kivelson, and Gergely Zimanyi.
\newblock {\em Phys. Rev. B}, 37(7):3283--3294, Mar 1988.

\bibitem{MPAF}
Matthew P.~A. Fisher.
\newblock {\em Phys. Rev. Lett.}, 57(7):885--888, Aug 1986.

\bibitem{MPAF1}
Matthew P.~A. Fisher.
\newblock {\em Phys. Rev. B}, 36(4):1917--1930, Aug 1987.

\bibitem{GIL}
Gil Refael, Eugene Demler, Yuval Oreg, and Daniel~S. Fisher.
\newblock {\em Physical Review B (Condensed Matter and Materials Physics)},
  75(1):014522, 2007.

\bibitem{aji1}
Vivek Aji and C.~M. Varma.
\newblock {\em Physical Review Letters}, 99(6):067003, 2007.

\bibitem{aji2}
Vivek Aji and C.~M. Varma.
\newblock {\em Physical Review B (Condensed Matter and Materials Physics)}, 79(18):184501, 2009.

\bibitem{aji3}
Vivek Aji, Arkady Shekhter and C.~M. Varma.
\newblock {\em Physical Review B (Condensed Matter and Materials Physics)}, 81(6):064515, 2010.

\bibitem{as1}
Arkady Shekhter and C.~M. Varma.
\newblock {\em Physical Review B (Condensed Matter and Materials Physics)}, 80(21):214501, 2009.

\bibitem{MW}
N.D. Mermin and H.~Wagner.
\newblock {\em Phys. Rev. Lett.}, 17:1113, 1966.

\bibitem{PCH}
P.C. Hohenberg.
\newblock {\em Phys. Rev.}, 158:383, 1967.

\bibitem{BER}
V.L. Berezinskii.
\newblock {\em Zh. Eksp. Teor. Fiz.}, 59:907, 1970.

\bibitem{KT}
J.M. Kosterlitz and D.J. Thouless.
\newblock {\em Jour. Phys. C}, 6:1181, 1973.

\end{thebibliography}
%
%

\end{document}